\documentclass[twocolumn,aps,prb]{revtex4}
\usepackage{graphics,graphicx}
\usepackage{dcolumn}
\usepackage{bm}

\newcommand{\Ref}[1]{Ref.~\onlinecite{#1}}

\def\eb{\begin{equation}}   
\def\ee{\end{equation}}     
\def\ea#1{\begin{eqnarray} #1 \end{eqnarray}}   

\def\shro{Schr\"odinger}

\def\ra{\rightarrow}

\def\im{\text{Im}}
\def\re{\text{Re}}

\def\of#1{\left(#1\right)}

\def\eq#1{Eq.~(\ref{#1})}
\def\eqs#1#2{Eqs.~(\ref{#1}) and (\ref{#2})}

\def\sof#1{\left[ {#1} \right]}

\def\Dlt{\Delta}

\def\Ppm{\Psi_\pm}
\def\Pmp{\Psi_\mp}

\def\sc{{\text{sc}}}

\def\apm{\alpha_\pm}
\begin{document}

\author{Corey Trahan and Bill Poirier}
\affiliation{Department of Chemistry and Biochemistry, and
         Department of Physics, \\
          Texas Tech University, Box 41061,
         Lubbock, Texas 79409-1061}
\email{Bill.Poirier@ttu.edu}

\title{Reconciling Semiclassical and Bohmian Mechanics: \\
III. Scattering states for continuous potentials}

\begin{abstract}

In a previous paper [J. Chem. Phys. {\bf 121} 4501 (2004)]
a unique bipolar decomposition, $\Psi = \Psi_1 + \Psi_2$ was
presented for stationary bound states $\Psi$ of
the one-dimensional \shro\ equation, such that the components
$\Psi_1$ and $\Psi_2$ approach their semiclassical WKB analogs in
the large action limit. The corresponding bipolar
quantum trajectories, as defined in the usual Bohmian mechanical
formulation, are classical-like and well-behaved,
even when $\Psi$ has many nodes, or is wildly oscillatory.
A modification for discontinuous potential stationary
stattering states was presented in a second paper
[J. Chem. Phys. {\bf !!!} !!!! (2005)], whose generalization for
continuous potentials is given here. The result is an
{\em exact} quantum scattering methodology using {\em classical}
trajectories. For additional convenience in handling the tunneling
case, a constant velocity trajectory version is also developed.

\end{abstract}

\maketitle


\section{INTRODUCTION}
\label{intro}

This paper is the third in a series investigating the
use of ``counter-propagating wave methods''
(CPWMs)\cite{poirier04bohmI,poirier05bohmII,babyuk04,wyatt} for
solving the \shro\ equation exactly. CPWMs are a particular
variant of the more general multipolar decomposition methods,
wherein the wavefunction $\Psi$ is decomposed into two or
more components, $\Psi_j$. Thus, for a two-term, or ``bipolar''
decomposition, $\Psi = \Psi_1 + \Psi_2$.  This rather trivial-seeming
procedure can be quite advantageous in the context of
quantum trajectory methods
(QTMs),\cite{wyatt,lopreore99,mayor99,wyatt99,shalashilin00,wyatt01b,wyatt01c,burghardt01b,bittner02b,hughes03}
i.e. trajectory-based numerical techniques for performing exact quantum dynamics
calculations, based on Bohmian
mechanics.\cite{madelung26,vanvleck28,bohm52a,bohm52b,takabayasi54,holland}
Conventional QTMs use a single-term
or ``unipolar'' representation of the wavefunction from which
all other quantities, such as the quantum trajectories themselves,
are uniquely determined. Multipolar decomposition, on the
other hand, can lead to radically different QTM behavior
for the individual $\Psi_j$ components,
owing to the fact that the Bohmian equations of motion are
non-linear.\cite{poirier04bohmI,poirier05bohmII}

As applied to wavepacket dynamics for reactive scattering systems,
QTMs suffer from a significant and well-known numerical drawback, which
to date, precludes a completely robust application of these methods.
Namely, QTMs are numerically unstable in the vicinity of amplitude
nodes and ``quasi-nodes''
(i.e. rapid oscillations),\cite{wyatt,wyatt01b}
owing to singularities in the ``quantum potential,'' $Q$, which together
with the classical potential, $V$, determines the quantum trajectories.
In the reactive scattering context, such behavior is always observed,
due to interference between the incident and reflected waves. On
the other hand, if the latter two contributions to the total $\Psi$ were
somehow separated, and associated with two different interference-free
$\Psi_j$ components, the node problem might well be circumvented. If,
in addition, the $\Psi_j$ component field functions were smooth
and slowly-varying, far fewer QTM trajectories and time-steps might
be required than for $\Psi$ itself---although the latter could be
reconstructed at any desired time simply via linear superposition
of the components. These numerical advantages thus provide significant
motivation for consideration of the bipolar approach. A much more
detailed discussion may be found in the first two articles of this
series, paper~I (\Ref{poirier04bohmI}) and
paper~II (\Ref{poirier05bohmII}).

The most obvious aspect of any bipolar decomposition, including those
restricted along the lines of the preceding
paragraph,\cite{poirier04bohmI,poirier05bohmII,floyd94,brown02}
is that it is not unique.
The covering function method,\cite{babyuk04,wyatt} for instance,
treats $\Psi$ as the difference between two very large-amplitude
components, thus ``diluting'' the effects of interference. The
one-dimensional (1D) CPWM approach,\cite{poirier04bohmI,poirier05bohmII}
on the other hand, regards the bipolar decomposition
\eb
     \Psi = \Psi_+ + \Psi_- \label{psitot}
\ee
as a superposition of right- and left-traveling
counter-propagating waves, $\Ppm$. For stationary states at least,
the \eq{psitot} decomposition is defined such that the $\Ppm$
components correspond to semiclassical WKB approximations,
$\Ppm^\sc$, in the large-action limit. In addition to providing
pedagogical value (semiclassical and Bohmian mechanics can
not be so reconciled in a unipolar context), the semiclassical
field functions are typically smooth and slowly-varying, i.e.
the semiclassical-like CPWM components $\Ppm$ provide the
desirable numerical advantages described in the preceding paragraph.

In paper~I (\Ref{poirier04bohmI}), a {\em unique} CPWM
bipolar decomposition was determined for bound stationary
eigenstates of arbitrary 1D Hamiltonians. The resultant $\Ppm$
field functions are smooth and interference-free, and approach
the WKB approximations in the large-action limit
(within the classically allowed region of space) as desired.
Moreover, the quantum potentials $q_\pm$ become vanishingly
small in this limit, so that the bipolar quantum trajectories
approach classical trajectories, and only a small number
are required for a numerical propagation, regardless
of excitation energy, $E$. In contrast, the unipolar
$\Psi$ exhibits {\em motionless} trajectories, and an
arbitrarily increasing number of nodes in the
large-action/energy limit. Results were presented for both the
harmonic and Morse oscillator potentials.

In paper~II (\Ref{poirier05bohmII}), the 1D CPWM ideas were modified
somewhat for stationary scattering states of discontinuous potentials.
Although the CPWM decomposition of paper~I is uniquely specified
for any arbitrary 1D eigenstate---bound or scattering---and in the
bound case, always satisfies the correspondence principle, the
non-$L^2$ nature of the scattering states is such that the paper~I
decomposition generally does {\em not} satisfy correspondence
globally. As discussed in Sec.~\ref{semiclassical}, this requires
that global modifications must be made in order
to enable a correspondence between $\Ppm^\sc$ and $\Ppm$. These
are such as to lead to substantial differences between the
density {\em functions} $|\Ppm^\sc(x)|^2$ and $|\Ppm(x)|^2$, although
the {\em trajectories} are identical. Moreover, the resultant $\Ppm$
are found to correspond to the familiar ``incident,'' ``transmitted,''
and ``reflected'' waves of traditional scattering theory\cite{taylor}
in the appropriate asymptotic limits. From the time-dependent standpoint,
reflection was found to be due to trajectory {\em hopping} from
one CPWM component to the other, as can be naturally understood
using a simple ray optics analogy. The method was applied to
several elementary discontinuous potential systems, including
the square barrier/well.

In the present paper (paper~III), the \Ref{poirier05bohmII} formulations
are generalized to incorporate both continuous {\em and} discontinuous
potential systems. Once again, an analogy is drawn from semiclassical
mechanics, albeit a ``sophisticated'' version\cite{heading,froman,berry72}
less frequently considered (Sec.~\ref{semiclassical}). As in paper~II,
a time-dependent method based on ray optics is developed for
computing stationary states of any desired energy and boundary
conditions. For the most part, the discussion of the preceding
paragraph still applies, but some additional key points
should be emphasized. First, the bipolar
decomposition now provides a sensible definition of ``incident,''
``transmitted,'' and ``reflected'' waves {\em throughout all space},
not just asymptotically. Second,
the explicit hopping of trajectories from one CPWM component to
the other is replaced with a {\em coupling term} in the
time-evolution equations. Third, the trajectories become
{\em completely classical}.
Finally, an alternative methodology is also developed, based on
the use of {\em constant velocity trajectories}, which can be
readily applied to barrier tunneling situations.
The new methods are found to be remarkably efficient, accurate,
and robust across a diverse range of 1D test potentials
and system energies (Sec.~\ref{results}).

The paper is organized as follows. A derivation and discussion
of the time-evolution equations for the CPWM bipolar components
$\Ppm$, both for classical and constant velocity trajectories, are
presented in Sec.~\ref{theory} and the Appendices.
Sec.~\ref{numerical} provides numerical details of the various
bipolar algorithms used to compute stationary states. Results
are presented in Sec.~\ref{results} for four benchmark applications:
Eckart barrier; square barrier; uphill ramp; double-Gaussian barrier.
For the first three, these are compared with known analytic solutions.
Concluding remarks, including prospects for future development,
may be found in Sec.~\ref{conclusion}.


\section{THEORY}
\label{theory}

\subsection{Semiclassical Approximations}
\label{semiclassical}

Let $V(x)$ be the potential energy for a 1D scattering system,
and $E$ the energy of some stationary state $\Psi(x)$ such that
$E > V(x)$ for all $x$. In reality, $\Psi(x)$ always manifests
some reflection, even though the energy is above the potential
barrier. However, basic WKB theory predicts {\em zero} reflection
in this case, as the classical trajectories do not turn around.
This is also evident from the form of the two basic WKB
counter-propagating wave solutions,
\eb
\Ppm^\sc(x) = r_\sc(x) e^{\pm i s_\sc(x)/\hbar}, \label{scsoln}
\ee
where
\eb
r_\sc(x) = \sqrt{{m F \over s'_\sc(x)}} \quad\text{and}
\quad s'_\sc(x) = \sqrt{2 m \sof{E-V(x)}}, \label{scrs}
\ee
$m$ is the mass, $F$ is the invariant flux (paper~I),
and primes denote spatial differentiation.
Note that since scattering solutions are non-square-integrable,
the choice of $F$ is arbitrary; however, throughout this
paper we shall adopt the usual left-incident wave
normalization convention,
$\lim_{x\ra -\infty} r_\sc(x) = 1$, so that
$F=\lim_{x \ra -\infty} s'_\sc(x)/m$. The positive and negative momentum
functions, $p_\pm^\sc(x) = \pm s'_\sc(x)$, specify the classical trajectories and
semiclassical
Lagrangian manifolds\cite{poirier04bohmI,keller60,maslov,littlejohn92} (LMs)
associated with the $\Ppm^\sc(x)$ solutions.
Clearly, the classical trajectories do not change direction, as there
are no turning points along the real $x$ axis. Therefore, for an arbitrary
linear combination solution,
\eb
\Psi = \alpha_+ \Psi_+^\sc + \alpha_- \Psi_-^\sc, \label{alphaeq}
\ee
the entire reflected wave must be due to the $\Psi_-^\sc(x)$ contribution.
The boundary conditions presume that the left-traveling wave
contribution vanish in the $x\ra \infty$ limit, implying that
$\alpha_- = 0$ and $\Psi(x) = \Psi_+(x)$---i.e., there {\em is} no
reflected wave.

Various strategies have been developed to deal with the
above difficulty of the basic WKB
method.\cite{heading,froman,berry72} The most
common involve analytic continuation into the complex plane.
Specifically, if no turning point is located along the real
axis, then one is found elsewhere in the complex plane. The
path of the incident wave is deformed so as to give rise to
a reflected wave upon encountering the complex turning point.
An analysis of the Stokes and anti-Stokes
lines\cite{heading,froman,berry72,child,poirier03capII}
that emanate from the complex turning
point and cross the real $x$ axis, enables one to
effectively recast the above procedure in terms of purely real-valued $x$.
The net effect (as interpreted in this paper)
is the introduction of {\em coupling} from $\Psi_+ = \alpha_+ \Psi_+^\sc$
to $\Psi_- = \alpha_- \Psi_-^\sc$, resulting
in an $\alpha_-$ value that {\em changes} over $x$,
thus yielding meaningful semiclassical partial reflection probabilities.

Alternatively, there is an approach due to
Bremmer\cite{heading,berry72,bremmer51}
that from the start is formulated entirely on the real $x$ axis.
This approach is preferred for the present purpose, as it provides
a common bipolar foundation not only for basic WKB and
``sophisticated'' (i.e. capable of predicting partial reflection)
semiclassical methods, but also for {\em exact quantum} scattering applications,
as will be shown in Sec.~\ref{classicaltraj}. Bremmer's idea is to
model the continuous potential $V(x)$ using a collection of
discontinuous steps. Solutions are determined for an arbitrary
step size, which is then made infinitesimally small.

Locally, along any given step, the two ``exact'' \shro\ solutions
(for the model step potential) are plane waves,
$\exp(\pm i p_k x)$, where $p_k=\sqrt{2 m \of{E-V_k}}$, and $V_k$
is the (constant) potential value along the $k$'th step (Appendix~A).
A viable global solution, $\Psi(x)$, must match boundary conditions
appropriately at each of the step edges. In reality, a pure positive
momentum local plane wave in one step would be joined to some
linear combination of positive and negative momentum plane waves in
the adjacent step, corresponding to partial reflection and transmission
off of the local step. If the reflection contribution is {\em ignored},
e.g. so that only {\em positive} momentum local plane waves are involved,
then the resultant approximate global solution can be easily shown to
be just $\Psi_+^\sc(x)$ in the limit of infinitesimal step size. Similarly,
$\Psi_-^\sc(x)$ is obtained from the negative momentum local plane
waves. This approximation thus leads to the uncoupled basic WKB
solutions of \eqs{scsoln}{scrs}.

The above approach clearly demonstrates how
``continuous reflection''\cite{poirier03capI}---i.e. that arising from
continuous potentials---may be interpreted in more familiar discontinuous
terms. It also suggests that {\em reflection requires coupling between
positive and negative momentum wavefunction components}. This approach
has been used to develop a sophisticated WKB approximation, similar to the
Stokes/anti-Stokes approach described above, wherein one-way reflection
from $\Psi_+$ to $\Psi_-$ is retained, but back reflection is
ignored.\cite{berry72} On the other hand, if {\em all} reflection
is retained, one can derive a coupled pair of first order differential
equations for $\Ppm$ that {\em exactly} describes the quantum scattering
solution $\Psi$ of \eq{psitot}. Although the $\Ppm$ components are
not themselves stationary solutions, they do correspond to the familiar
incident/transmitted/reflected interpretations as discussed in paper~II.

\subsection{Exact Quantum Dynamics Using Classical Trajectories}
\label{classicaltraj}

\subsubsection{Time-evolution equations}

In paper~II, an exact quantum, CPWM bipolar decomposition scheme for stationary
scattering states was presented such that a suitable $\Ppm$ could be constructed
for a solution $\Psi$ with any desired boundary conditions,
using a simple time-dependent approach.
The classical trajectories and LMs were found to be identical
to those of the basic WKB approximations, $\Ppm^\sc(x)$---thus satisfying
the correspondence principle, and giving rise to smooth, well-behaved,
interference-free CPWM component functions, $\Ppm$. Although the method formally
applies only to discontinuous potentials, Sec.~\ref{semiclassical} suggests
that it ought to be extendible to continuous potentials as well, simply
by modeling the latter using infinitesimally small steps.
This idea is indeed quite straightforward to implement, as discussed in
Appendix~A and Fig.~\ref{derivefig}. As in paper~II, the result is a
trajectory-based scattering methodology that is both {\em local} and {\em exact}
in its treatment of reflection and transmission, due to the
time-dependent nature of the approach. In contrast, the global
character of reflection and transmission in conventional time-independent
exact quantum methods is well-recognized.  Of the
time-independent semiclassical approximations, the most sophisticated
depend on the relative placement of local potential features such as
discontinuities and turning points,\cite{heading,froman,berry72}
although the more approximate methods treat such features independently.

\begin{figure}
\includegraphics[scale=0.65]{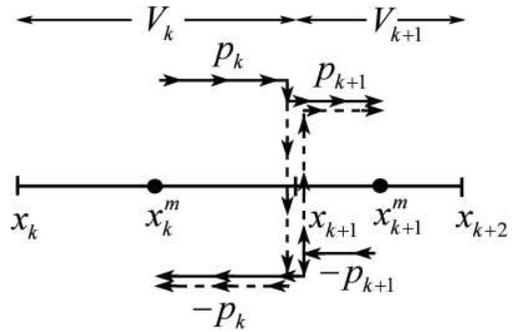}
        \caption{Time-dependent analysis of the $\Ppm$ coupling, obtained
                 via trajectory hopping due to local reflection off of
                 infinitesimal potential steps. Continuous potential reflection
                 is thus described in more pedagogical discontinuous terms.}
        \label{derivefig}
\end{figure}

From Appendix~A, the hydrodynamic (Lagrangian) time-evolution equations for the
CPWM bipolar components are found to be
\eb
     {d \Ppm \over dt} = \sof{ \mp {p' \over 2m} + {i \over \hbar} \of{E-2V}} \Ppm
     \pm {p' \over 2m} \Pmp, \label{Pdotgwf}
\ee
where $\pm p = \pm \sqrt{2 m (E-V)}$ are the momenta, which
define the $\Ppm$ LM and trajectories. Note that as expected, these are
identical to those of $\Ppm^\sc(x)$, i.e.
{\em the trajectories are completely classical} even though the solutions
are exactly quantum mechanical (Appendix~B). All of the other desirable
properties, as described in the preceding paragraph and in paper~II,
are also found to be true.

\subsubsection{Comparison with unipolar Bohmian mechanics}

\label{compare}

It is worthwhile to compare \eq{Pdotgwf} with the usual unipolar evolution
equations of Bohmian mechanics.\cite{wyatt,bohm52a,bohm52b,holland}
Both are exact QTMs, but quantum
effects are incorporated in very different ways. In unipolar Bohmian mechanics,
the trajectory evolution is determined by the modified potential $V+Q$, where
the quantum potential correction $Q$ is responsible for all quantum effects.
Evaluation of $Q$ requires explicit double spatial differentiation
of the wavefunction $\Psi$, which in turn requires specialized numerical
techniques (Sec.~\ref{lowlevel}). Also, $Q$ diverges at nodes,
and is otherwise numerically unstable in the presence of interference.
In contrast, the present bipolar QTM scheme avoids interference difficulties
as desired, owing to the semiclassical correspondence. Quantum effects do
{\em not} manifest in the trajectories themselves (as these evolve completely
classically), and there {\em is} no quantum potential in \eq{Pdotgwf}.

So where do quantum effects come from in \eq{Pdotgwf}? As anticipated
in Sec.~\ref{semiclassical}, these must be due to inter-component coupling,
i.e. the last term in the equation. Note that the amount of coupling is
proportional to $p' \propto V'$. Thus, the coupling vanishes in the limit
that $V' \ra 0$ which
is reasonable, considering that this is also the usual WKB condition
and that the basic WKB solutions are uncoupled. Note
that {\em no} wavefunction spatial derivatives are required in
\eq{Pdotgwf}---a decided advantage over the unipolar approach.
In principle, every trajectory ``splits'' into a forward and backward
moving trajectory pair at every point in space and time (paper~II).
The forward-moving
trajectory remains on the same LM as the source, whereas the backward trajectory
``hops'' onto the other LM. However, unlike the situation in paper~II,
this splitting, hopping, and subsequent recombining need not be considered
explicitly, as it is all implicitly dealt with at the differential equation
level. Nevertheless, conceptually, one may regard trajectory hopping as
the {\em source} of coupling and quantum effects.

Note that for scattering systems, $V'\ra 0$ in the asymptotic
limits of $x$. This implies that there is no asymptotic coupling (or
trajectory hopping) between the two $\Ppm$ components---an essential
feature from the perspective of computing scattering quantities
such as global reflection and transmission probabilities (which according
to normalization and boundary conventions discussed in Sec.~\ref{semiclassical},
are obtained respectively via $P_{\text{refl}} = \lim_{x\ra-\infty} |\Psi_-(x)|^2$
and $P_{\text{trans}} = \lim_{x\ra\infty} \sof{p(x)/p(-x)}|\Psi_+(x)|^2$).
Equation~(\ref{Pdotgwf})
also ensures that the asymptotic $\Ppm(x,t)$ solutions
are the desired {\em plane waves}, and not some arbitrary linear
superposition such as a sine wave. Note that in this regard, the
hydrodynamic time-derivative aspect of \eq{Pdotgwf} is {\em essential},
e.g. the ordinary \shro\ equation would not preclude asymptotic
sine wave solutions.

Although $\Ppm$ and $\Ppm^\sc(x)$ are identical in the
asymptotic limits (apart from the constant scaling factor $\apm$),
they differ in the interaction region [i.e. $\apm(x)$ depends on $x$
in this region], even though the trajectories and LMs are the same throughout.
This raises some interesting questions vis-a-vis the interpretation of
standard Bohmian quantities in a coupled bipolar context, which will be
explored more fully in Sec.~\ref{additional}. Here, we consider the
bipolar quantum potential, which in the conventional sense would be
obtained via $q_\pm = -(\hbar^2/2m) (r_\pm''/r_\pm)$ with
$\Ppm = r_\pm \exp(i s_\pm/\hbar)$ and $r_\pm$ and $s_\pm$ real. But,
it is not clear that such a definition should apply in the case where
the $\Ppm$ are coupled and do not individually evolve according to the
\shro\ equation. Indeed, such a definition would be inconsistent with
the time evolution of $s_\pm'$, which in any event is itself inconsistent
with the classical trajectory evolution ($s_\pm' \ne \pm p$). In this
context, it is perhaps more natural to define $q_\pm$ such that
$(V+q_\pm)$ determines the trajectory evolution. According
to this definition, $q_\pm=0$ for the present
bipolar CPWM formulation, even throughout the interaction region.

\subsubsection{Additional properties}

\label{additional}

As in paper~II, the desired stationary state solution, as obtained from
the \eq{Pdotgwf} evolution equations, is not observed at all times,
but only asymptotically in the large $t$ limit. The same ray optics
and continuous wave cavity ring-down interpretations that apply in
paper~II also apply here. Thus, at $t=0$, one starts with a left-incident
asymptotic plane wave truncated outside the interaction region.

It can be shown (paper~II, Appendix~B) that as $t\ra\infty$, within
any finite $x$ interval that includes the interaction potential, the resultant
$\Psi(x,t)$ converges exponentially quickly to the correct time-dependent
stationary state solution with appropriate $x$ boundary conditions.
A proof is provided in Appendix~B, which, for comparison with the
time-dependent \shro\ equation, relies on the Eulerian
(partial time derivative) version of \eq{Pdotgwf}, i.e.
\eb
     {\partial \Ppm \over \partial t} = \mp {p \over m} \Ppm' +
            \sof{ \mp {p' \over 2m} + {i \over \hbar} \of{E-2V}} \Ppm
            \pm {p' \over 2m} \Pmp \label{Pdot}.
\ee
Note that
\eq{Pdot} involves a single spatial derivative of the wavefunction,
unlike \eq{Pdotgwf}.

\begin{figure}
\includegraphics[scale=0.5]{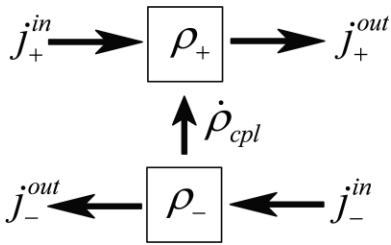}
        \caption{Schematic indicating the flux, probability, and continuity
                 relationships that exist between $\Ppm$ components for the
                 CPWM bipolar decompositions considered in this paper.
                 All probability that leaves the $\Psi_-$ component must flow into
                 the $\Psi_+$ component. Consequently,
                 $\int \sof{\rho_+(x) + \rho_-(x)}\,dx$ is conserved for all time.}
        \label{fluxfig}
\end{figure}

Equation~(\ref{Pdot}) above can be employed to derive a very useful
flux relationship,
\eb
     {\partial \rho_\pm \over \partial t} = - j_\pm' \pm
       {p' \over m} \re \sof{{\Psi_+}^* \,\Psi_-}, \label{fluxrel}
\ee
where $\rho_\pm = |\Ppm|^2$ is the density, and $j_\pm = \pm (p/m) \rho_\pm$
is the flux, defined in terms of the actual classical trajectory
velocities. Taken individually, the $\partial \rho_\pm / \partial t$
do not obey continuity (except when $p'=0$),
implying that the total probability for each $\Ppm$ component is
not conserved.  Together, however, they do satisfy a kind of
continuity relation, in that
$\partial (\rho_+ + \rho_-) / \partial t = - (j_+ + j_-)'$,
implying that the {\em total probability for both $\Psi_+$ and
$\Psi_-$ is conserved}. This is an important, nontrivial
result---quite distinct from the usual probability conservation
of $\Psi$ itself. As described in
Fig.~\ref{fluxfig}, \eq{fluxrel} in effect states that
$\partial \rho_\pm/ \partial t$ is equal to the usual negative
flux divergence {\em plus} a coupling term,
$\pm \partial \rho_{\text{cpl}}/ \partial t =
\pm (p'/m) \re \sof{{\Psi_+}^*\, \Psi_-}$,
representing the rate at which probability flows from one
CPWM component to the other.

It should be stated that the classical trajectory CPWM bipolar decomposition
scheme, in the form described above, is not viable for computing
stationary eigenstates below the potential barrier maximum.
Tunneling {\em per se} is not the
issue, since one can in principle apply \eq{Pdotgwf} in the tunneling regime
using analytic continuation (in a manner similar to that applied in
paper~II) as has been confirmed in numerical tests.
However, a difficulty arises when
$|p|\ra 0$ and $|p'| \ra \infty$, i.e. in the vicinity of the real $x$
axis turning points. For discontinuous potentials, this only occurs
for energies near a piece-wise barrier energy $V_k$,
manifesting as substantially increased propagation times (paper~II).
For continuous potentials, all energies below the barrier height exhibit
turning points along the real $x$ axis, and are therefore problematic.
Specifically, the $p'$ terms in \eqs{Pdotgwf}{Pdot} lead to numerical
instabilities near the turning points.

\subsection{Exact Quantum Dynamics Using Constant Velocity Trajectories}
\label{constanttraj}

\subsubsection{Motivation}
\label{motivation}

It should be noted that turning point/caustic issues similar
to those described in Sec.~\ref{additional} are also faced by
semiclassical methods. Thus, similar remedies may presumably
be applied here (e.g. complex plane path deformation),
although these are not
considered further in this paper.  Instead, we are guided by
one of our original motivations,\cite{poirier04bohmI,poirier05bohmII}
to develop methods that avoid such semiclassical
difficulties altogether. To this end, we turn once again to
the ray optics interpretation of the bipolar approach,
as discussed in paper~II.

Note that in classical optics, the ``ray'' interpretation of a given
time-evolving electromagnetic field is not necessarily
unique; one has a certain freedom to define rays
as convenient for a given application.\cite{jackson}
Consider the case of total internal reflection, for instance,
at the boundary between two materials with different indices
of refraction. Quantum mechanically, this corresponds
to a single-step discontinuous potential at an energy below the
barrier height (paper~II). According to the usual ray interpretation,
the incident rays refract to the extent that they become parallel
to the interface, and therefore do not penetrate at all into the
second medium. This picture is physically somewhat incorrect however,
in that it does not capture the exponentially-damped evanescent
wave.\cite{jackson,brillouin14}
Quantum mechanically, this corresponds to tunneling into the
discontinuous step, which is simply not described using the bipolar
classical trajectory method as presented in Sec.~\ref{classicaltraj}.

On the other hand, a simple, alternative ray interpretation can be
applied to total internal reflection that {\em does} predict the
evanescent wave correctly. According to this interpretation,
{\em the incident ray velocities remain} constant
{\em across the interface}. These rays are reflected
within the second medium, giving rise to the evanescent wave,
and also providing an explanation for the
Goos-H\"anchen phenomenon.\cite{poirier05bohmII,jackson,hirschfelder74}
The quantum-mechanical analog would be {\em constant velocity}
bipolar trajectories, for which the incident wave
asymptotic velocities remain constant {\em throughout}
the interaction region. Presumably, the time
evolution equations derived from such a choice of trajectories would
not depend very sensitively on the energy $E$, even in the vicinity of
the barrier height, so that tunneling, and classical
turning points/caustics would pose no special difficulties.

\subsubsection{Fr\"oman and Fr\"oman methodology}

The semiclassical-like CPWM discussed in Sec.~\ref{classicaltraj} and
the Appendices does not generalize in any straightforward manner for
trajectories other than classical. However, there is an alternative
time-independent formulation, conceptually similar to the Bremmer
(B) approach but differing in the details, which does allow for such a
generalization. This approach is due to Fr\"oman and Fr\"oman
(F),\cite{froman} who use it to define generalized semiclassical
approximations, although it can also be used to derive
corresponding exact quantum bipolar decompositions. The guiding
principle of the F approach (as interpreted here) is that
semiclassical solutions should satisfy the invariant flux
property\cite{poirier04bohmI,poirier05bohmII,froman}---i.e., the left side of
\eq{scrs} with $s'_\sc(x)$ essentially arbitrary.
If $s'_\sc = p$ is chosen classically, then the usual
basic WKB solutions result for $\Ppm^\sc(x)$.  Curiously, however,
the corresponding  exact quantum $\Ppm(x)$ are {\em not} the same as in
Sec.~\ref{classicaltraj}. In the F case, the decomposition of
\eqs{psitot}{alphaeq} is uniquely defined via the time-independent
\shro\ equation and the relation
\ea{
     \Psi' & = & \alpha_+ {\of{\Psi_+^\sc}}' + \alpha_- \of{{\Psi_-^\sc}}'
          \label{FPprimeA} \\
           & = & -{p' \over 2p} \Psi +
          {i p \over  \hbar} \of{\Psi_+ - \Psi_-}. \label{FPprimeB}}
Thus, in comparing \eq{FPprimeA} to \eq{alphaeq}, the
$\apm$ coefficients behave in the derivative {\em as if}
they were $x$-independent constants, though, in fact, they are not
(except when there is no coupling).

Although not identical, the F and B bipolar decompositions are somewhat
similar in the classical trajectory case [compare \eq{FPprimeB} to \eq{BPprime}],
and both are plagued by similar numerical
instabilities near turning points, as has been confirmed in numerical testing.
However, the main advantage of the F approach is that it can be applied
to arbitrary trajectories, $p$. In particular, we choose
{\em constant velocity} trajectories obtained from the
left-incident plane waves, i.e. $p = \sqrt{2 m E}$
[with $\lim_{x \ra -\infty} V(x) = 0$].  This gives rise to exact
quantum CPWM components $\Ppm$ that are smooth and well-behaved
everywhere, even near turning points and barrier maxima. Note that
for constant velocity trajectories, \eqs{psitot}{FPprimeB}
imply that $\Ppm$ are linear combinations of $\Psi$ and $\Psi'$.

\subsubsection{Time-evolution equations}

In analogy with Sec.~\ref{classicaltraj}, the goal is to develop a
time-dependent method to compute the F constant velocity CPWM
bipolar decomposition in the large $t$ limit. However, since the
F construction of the $\Ppm$ is radically different from that of B,
a substantially different approach than that of Appendix~A must be used.
We have developed several different
derivations, all of which yield the same final results. The simplest strategy
is to ``work backwards'' through Appendix~B, but using \eq{FPprimeB} for
constant velocity trajectories instead of \eq{BPprime}. However, a more
pedagogical approach may also be employed, as presented below.

As in \eq{FPprimeA}, the basic idea is to presume that the coefficients
$\apm$ act as constants, but with respect to {\em time} derivatives
rather than spatial derivatives. In particular, it seems more natural
here to refer to {\em total} rather than partial time derivatives, given
the trajectory-based nature of the methodology. Less obvious at this stage
is the fact that the linear combination $\of{\Psi_+ - \Psi_-}$ must be
used rather than $\Psi = \Psi_+ + \Psi_-$ itself, in order to be consistent
with the time-independent F results. We thus obtain the condition
\ea{
     {d \Psi_+ \over dt} -  {d \Psi_- \over dt} & = &
     \alpha_+ {d \Psi_+^\sc \over dt} - \alpha_+ {d \Psi_+^\sc \over dt}
     \nonumber \\
     & = & - {i E \over \hbar}\of{\Psi_+ - \Psi_-} + {p \over m} \Psi'
     \label{Ffirstcond} }
Together with the time-dependent \shro\ equation
(converted to total derivative form), as applied
to the \eq{psitot} linear combination,
\eq{Ffirstcond} above gives rise to a unique set of
time-evolution equations for $d \Ppm / dt$. In the constant velocity
trajectory case, for which
$\Ppm^\sc(x,t) \propto \exp\sof{i (\pm \sqrt{2 m E} x - E t)/\hbar}$,
these are found to be
\eb
     {d \Ppm \over dt} = {i \over \hbar} \of{E-V}\Ppm -
                         {i \over \hbar} V \Pmp \label{Pdotunigwf}.
\ee

\subsubsection{Additional properties}
\label{additionaluni}

Using a procedure analogous to that described at the start of
Appendix~B, one can show that for stationary states, \eq{Pdotunigwf}
is in fact consistent with the time-independent F constant velocity
CPWM bipolar decomposition.
This requires the Eulerian version of \eq{Pdotunigwf},
i.e.
\eb
     {\partial \Ppm \over \partial t} = \mp {p \over m} \Ppm' +
                         {i \over \hbar} \of{E-V}\Ppm -
                         {i \over \hbar} V \Pmp \label{Pdotuni}.
\ee
What is not so clear, however, is whether starting
with a truncated asymptotic plane wave, one necessarily
approaches the stationary solution in the large $t$ limit
(Sec.~\ref{additional}, paper~II). Although this has not yet been
proven, it is a reasonable assumption, given both
the counter-propagating trajectory nature of the method
and flux properties similar to the B classical trajectory case
(discussed below).
Moreover, for all of the numerical applications considered in
Sec.~\ref{results}, exponentially fast convergence is in fact observed.

The F constant velocity
time-evolution equations [\eq{Pdotunigwf}] offer some decided
advantages over the classical trajectory approach. Since only
energy quantities appear on the right hand side, there is no
need to resort to analytic continuation in order to handle
tunneling for the below-barrier energies.
The equations may therefore be applied
with equal ease throughout the energy spectrum, and in fact, the
resultant $\Ppm$ are qualitatively similar above, below, and just
at the potential barrier maximum (Sec.~\ref{eckart}).
Another advantage is
that the numerical propagation does not require spatial
differentiation of any kind---not even of the potential, V,
to determine forces driving the trajectories.
Furthermore, \eq{Pdotunigwf} may be
numerically implemented as is for discontinuous potentials,
just as easily as for continuous potentials, without the need
for explicit splitting and recombining of trajectories as
in paper~II (Sec.~\ref{squarebarrier}).

On the other hand, the F constant velocity approach
introduces a drawback that the classical trajectory methods
do not have to contend with when the potential is
``asympotically asymmetric''---by which we mean simply that
$\lim_{x \ra -\infty} V(x) = 0 \ne \lim_{x \ra \infty} V(x)= V_0$,
corresponding e.g. to an exoergic or endoergic chemical reaction.
Whereas in the $x\ra -\infty$ limit, the evolution equations
are uncoupled, this is not true in the $x \ra \infty$ limit
if $V_0\ne0$, in which case the time-dependent $\Ppm$
are not expected to converge. Many techniques could
be used to remedy this situation, e.g. trajectories described
via smoothly-varying sigmoid (tanh-like) functions rather
than uniform or classical trajectories. For purposes of
this paper, we adopt a simpler solution, wherein {\em two}
F constant velocity CPWM bipolar decompositions are used, one each
for the reactant and product regions. Standard boundary
condition matching techniques are then applied to join these
together (Sec.~\ref{asymmetric}).

Regarding flux properties, \eq{Pdotuni} can be used to obtain
\eb
     {\partial \rho_\pm \over \partial t} = - j_\pm' \pm
       {2V \over \hbar} \im \sof{{\Psi_+}^* \,\Psi_-}, \label{fluxreluni}
\ee
which should be compared with \eq{fluxrel} for the B classical
trajectory case.
Note that the total combined probability for both $\Psi_+$ and
$\Psi_-$ is conserved here as well, i.e.
$\partial (\rho_+ + \rho_-) / \partial t = - (j_+ + j_-)'$,
and Fig.~\ref{fluxfig} still applies.  The coupling term
$\partial \rho_{\text{cpl}}/ \partial t =
(2V/\hbar) \im \sof{{\Psi_+}^*\, \Psi_-}$ is of course different
from that of Sec.~\ref{additional}, although in both cases, the
wavefunction inner product cross terms are involved. Several additional
relations unique to the constant velocity case may also be easily
derived from the above equations, e.g.
\eb
     d \rho_\pm / dt = \pm \partial \rho_{\text{cpl}}/ \partial t,
\ee
which states that the probability lost by a positive momentum trajectory
is gained by the negative momentum trajectory directly ``beneath'' it
(same $x$), and
\eb
     \rho_+' = \rho_-' \quad \text{for stationary solutions},
\ee
which states that the positive and negative component density functions
are identical apart from a constant. Note that for asymptotically
symmetric potentials, and the normalization and boundary conditions
described in Secs.~\ref{semiclassical} and~\ref{compare},
this constant must be the transmission probability itself,
$P_{\text{trans}}$.


\section{Numerical Details}
\label{numerical}

In this section, we discuss the numerical details associated
with implementing the bipolar time-evolution equations of
Sec.~\ref{theory} as a practical and robust algorithm for computing
stationary scattering states. Although any boundary conditions
may be considered, our emphasis shall be on left-incident solutions,
for which $\Psi_- \ra 0$ as $x \ra \infty$. The region of interest
is defined via $x_L \leq x \leq x_R$, a region presumed to include
the entire potential interaction to the desired level of
accuracy. According to the time-dependent ray optics picture
developed in paper~II, at the initial time, $\Psi = \Psi_+$
is a left-incident plane wave truncated on the right at $x=x_L$.
This initial wave propagates into the interaction region,
wherein it is coupled to $\Psi_-$ and eventually reaches a
stationary state. Outside of the interaction region,
the $\Ppm$ coupling may be regarded as effectively zero.
Thus, one may compute reflection and transmission
probabilities, $P_{\text{refl}}$ and $P_{\text{trans}}$,
simply by evaluating $\Psi_+(x_R)$ and $\Psi_-(x_L)$
at sufficiently large times.

From a purely numerical point of view, the above scheme offers
many advantages---e.g., the ability to compute specific scattering
states without the need for complex
scaling\cite{complexscaling78,reinhardt82,ryaboy94} or complex absorbing
potentials\cite{poirier03capII,poirier03capI,jolicard85,seideman92a,riss93,muga04}
that would increase the coordinate range unnecessarily.
Moreover, the {\em density}
of grid points may be substantially reduced,
owing to the fact that the interference-free
component functions $\Ppm$ are generally smooth and slowly varying
as compared to $\Psi$ itself. For the same reason, a larger
time-step is also anticipated. Most importantly however,
since $q_\pm = 0$, there is no need to compute on-the-fly numerical
spatial derivatives. With regard to parallel computation, therefore,
there is no need for trajectories to communicate {\em within} a LM,
although the coupling requires position-dependent communication
between positive and negative LM trajectories.

\subsection{Algorithmic Issues}
\label{algorithmic}

In order to solve the time-evolution equations
numerically, two trajectory grids are created, one for
each bipolar component of the total wavefunction.
Hereafter, ``upper'' will be used to describe attributes
of the $\Psi_+$ component, and ``lower'' for the $\Psi_-$
component (e.g. ``lower/upper grid,'' etc.). On each grid,
the corresponding wavefunction component is
spatially discretized at the grid point locations
and evolved in time.

For the applications considered here, we have found it
convenient to modify the ray optics picture somewhat
from the form described above and in paper~II.
First, we define an initial negative LM grid, even though
$\Psi_-$ itself is zero initially. The idea here is that
unlike paper~II, we wish to avoid explicit trajectory hopping.
Consequently, both sets of grid points exist for all time and
evolve independently of each other (though the component
wavefunctions do interact). Second, to avoid
unnecessary computational effort, no propagation is done
outside of the interaction region of interest. Instead,
at uniform time intervals, new $\Psi_+$ trajectories are fed
in at $x=x_L$, with an initial $\Psi_+$ value consistent
with the positive momentum asymptotic plane wave solution.
Later, as these upper grid trajectories reach $x= x_R$, they
are discarded. The $\Psi_-$ situation is similar, except that
the initial $\Psi_-(x_R)$ value is set to zero, and the
trajectories are discarded as they reach $x=x_L$.

The most substantial difference from the ray optics
picture, however, is found in our implementation of
the constant velocity method, for which the grids are
distributed uniformly throughout the interaction
region {\em at all times}. At the initial time, the upper
and lower grids coincide. The initial $\Psi_+$ is taken to be
the asymptotic plane wave solution {\em throughout}
the interaction region, and the initial $\Psi_-$
is set to zero.  The above modification, which we call the
``non-wavefront'' approach, is certainly not necessary, and
is introduced simply for convenience. Calculations
performed both ways reveal that both converge exponentially
to the correct stationary solution. However, the initial
convergence of the non-wavefront calculations is faster,
owing to the fact that the coupling takes effect immediately.
The numerical algorithm is also easier to implement.
Consequently, all constant velocity results presented
in Sec.~\ref{results} were obtained using the non-wavefront code.

As per Sec.~\ref{theory}, the upper and lower grids move
classically or with constant velocity, as appropriate, and the $\Ppm$
contributions evolve in accord with \eq{Pdotgwf} or \eq{Pdotunigwf}.
Since the grids move in opposite directions, they do not
remain commensurate over time. Numerical interpolation
(Sec.~\ref{lowlevel}) is therefore required to compute the
coupling contribution from the component wavefunction of one
grid to the other.  As an alternative to the above trajectory-based
(Lagrangian) approach, we have also implemented a fixed-grid
(Eulerian) bipolar propagation algorithm for the constant
velocity case.  The primary advantage of fixed-grid
propagation is that the two grids remain commensurate for all time,
thus avoiding the need for interpolation. However, this is achieved
at the expense of switching to the Eulerian evolution equations
[\eq{Pdotuni}], which require explicit numerical spatial
differentation of the bipolar wavefunction components.

\subsection{Low Level Issues}
\label{lowlevel}

For the numerical propagation of the time-evolution equations,
two fourth-order explicit integrators were considered---the multi-step
Adam's/Bashforth, and Runge-Kutta methods.\cite{phillips}
Although, both integrators are fourth-order accurate, and
require approximately the same CPU time, multi-step integrators
can not be trivially implemented for the asymptotically
asymmetric potential case (Sec.~\ref{asymmetric}).
Consequently, the basic fourth-order Runge-Kutta method
was used for all results presented herein. In the future,
the time integration can be made more accurate by using a
time-adaptive Runge-Kutta approach such as the Cash-Carp
method.\cite{press} For most of the calculations performed here,
a time-step of $\Dlt=10$ a.u. was found to be sufficient to achieve
a computed accuracy of $10^{-4}$ or better (Sec.~\ref{results}).
Generally speaking however, the time-step issue is quite problem-dependent,
as it can depend on both grid point velocities and spacing.

As discussed in Sec.~\ref{algorithmic}, the trajectory-based algorithms
require interpolation of both bipolar components in order to compute
coupling contributions. For all examples considered here, these
interpolates were calculated via a polynomial moving least squares
(MLS) method.\cite{wyatt} In MLS methods, local low-order polynomial
fits are calculated about each grid point via a small stencil of
surrounding neighbor points. To ensure that the fit will exactly
represent the function values at the grid point locations,
the stencil size and polynomial order must be set equal,
thus effectively transforming the MLS method into a
moving-interpolation method. For all calculations performed here,
five stencil points were used, corresponding to a symmetric stencil
(for the interior grid points) and quartic polynomial interpolation.

For the fixed-grid algorithms, explicit numerical spatial differentiation
of both CPWM bipolar components must be performed. For the present study,
these were calculated using centered, fourth-order finite
differences\cite{fornberg88} for the interior grid points, and one-sided
fourth-order finite difference for the boundary grid
points at $x_L$ and $x_R$.

\subsection{Asymptotically Asymmetric Potentials}
\label{asymmetric}

As discussed in Sec.~\ref{additionaluni}, the constant velocity
time-evolution equations will exhibit coupling in the
$x\ra\infty$ asymptote if $V(x)$ is asymptotically asymmetric,
i.e. if $\lim_{x\ra\infty}V(x) = V_0 \ne 0$. To remedy this,
we construct two sets of solutions, one for the ``reactant'' region,
$x \le x_0$, and another for the ``product'' region, $x \ge x_0$,
with $x_0$ the dividing point.
For the former solutions (denoted via the ``L'' subscript),
\eq{Pdotunigwf} may be used directly, with $p_L = \sqrt{2mE}$.
For the product solutions, however (denoted ``R''), the evolution
equations must be modified slightly to accommodate the
generalized asymptotic potential condition,
\eb
     {d \Psi_{R\pm} \over dt} = {i \over \hbar} \of{E-V-V_0}\Psi_{R\pm}
           -{i \over \hbar} \of{V-V_0} \Psi_{R\mp} \label{Pdotunigwfgen},
\ee
since the positive trajectory momentum is now
$p_R = \sqrt{2m(E-V_0)}$. It is clear from \eq{Pdotunigwfgen}
that the coupling vanishes as $x \ra \infty$.

For all propagation times, the wavefunction and its first derivative
must be continuous across $x_0$. Remarkably,
we can treat this boundary condition like an elementary plane
wave scattering off a step potential (paper~II).
This is achieved via \eq{FPprimeB}, in terms of which the two
exact conditions become
\ea{ \Psi_{L+}^0 + \Psi_{L-}^0 & = &
               \Psi_{R+}^0 + \Psi_{R-}^0 \nonumber \\
     p_L \sof{\Psi_{L+}^0 - \Psi_{L-}^0} & = &
     p_R \sof{\Psi_{R+}^0 - \Psi_{R-}^0},
     \label{asymbcA} }
where $\Psi_{L+}^0 =\Psi_{L+}(x_0)$, etc.
In the trajectory-based algorithm, the left and right
incident wavefunction values,
$\Psi_{L+}^0$ and $\Psi_{R-}^0$, are known at any given time,
whereas the locally transmitted values
$\Psi_{R+}^0$ and $\Psi_{L-}^0$ are unknown.
Equation~(\ref{asymbcA}) enables one to solve for the two unknowns,
and thus to continue propagating trajectories through the
dividing point, $x_0$.

The solutions are
\ea{
     \Psi_{R+}^0 &=& \of{{2 p_L \over p_L+ p_R}} \Psi_{L+}^0
         +\of{{p_R-p_L \over p_L+ p_R}} \Psi_{R-}^0 \nonumber \\
     \Psi_{L-}^0 &=& \of{{2 p_R \over p_L+ p_R}} \Psi_{R-}^0
         -\of{{p_R-p_L \over p_L+ p_R}} \Psi_{L+}^0, \label{asymbcB}}
which do indeed correspond to transmission and reflection
coefficients for stationary states of the discontinuous step
potential (Appendix~A and paper~II).
Numerically, the propagation is implemented as follows.
Equations~(\ref{Pdotunigwf}) and~(\ref{Pdotunigwfgen}) are used
until a trajectory reaches $x_0$, at which point it is replaced
with a new trajectory on the other side of $x_0$ via \eq{asymbcB}.
This requires a ``known'' value from the opposite LM. If the
grids are not commensurate (i.e. trajectories from opposite
LMs do not arrive at $x_0$ at the same time), it is necessary
to apply extrapolation to the opposite manifold to determine
the ``known'' value.


\section{RESULTS}
\label{results}

In this section, four different stationary scattering applications
are considered, all with comparable characteristic parameters---e.g.,
the same proton-like mass of  $m=2000$ a.u., barrier height
$V_0 = 400\,\text{cm}^{-1} \approx 0.0018$ a.u.
(note: $10\times$ smaller than in paper~II),
and interaction region, $[x_L=-3 \,\text{a.u.},\, x_R=3\, \text{a.u.}]$.
All four systems were solved using the non-wavefront, F, constant velocity
trajectory-based method, hereinafter referred to as the
``constant velocity trajectory'' scheme. For the Eckart system
(Sec.~\ref{eckart}), additional calculations were also performed
using the constant velocity fixed-grid scheme, and the wavefront, B,
classical trajectory method, now referred to as the ``classical
trajectory'' scheme.

\subsection{The Eckart Barrier}
\label{eckart}

\begin{figure}
\includegraphics[scale=0.8]{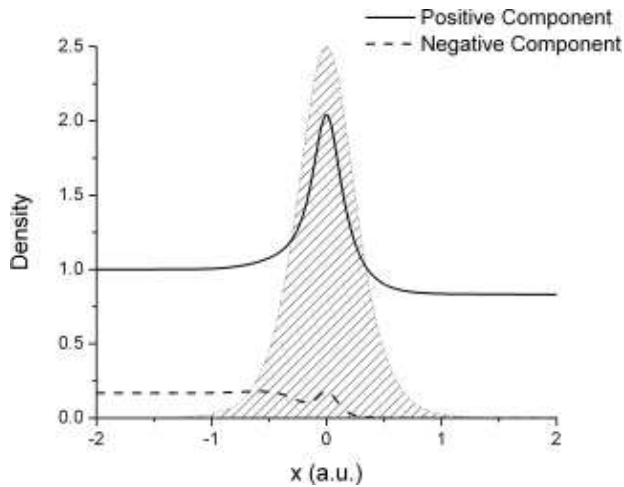}
        \caption{Converged positive component density $\rho_+(x)$ (solid line) and
                 negative component density $\rho_-(x)$ (dashed line) for the
                 $E=450\, \text{cm}^{-1}$ stationary state of the
                 $V_0=400\, \text{cm}^{-1}$ Eckart barrier system, as obtained
                 using the classical trajectory CPWM bipolar decomposition.
                 Shaded area indicates the Eckart potential $V(x)$.}
        \label{ebmanclassfig}
\end{figure}

The first application considered is the symmetric
Eckart problem,\cite{eckart30,ahmed93} defined via
\eb
     V(x)=V_0\,\text{sech}(\alpha x)^2, \label{eckartpot}
\ee
with parameter values $V_0=400\, \text{cm}^{-1}$,
and $\alpha=3.0$ a.u. The above potential is plotted on the
scale of the interaction region in Fig.~\ref{ebmanclassfig}.

We have performed numerical calculations for the Eckart system
using several of the previously described algorithms.
In the first study, the classical trajectory scheme was employed,
i.e. the continuous potential analog of paper~II. A time-step of
$\Dlt = 1$ a.u. was used, and a maximum of 45
trajectories per LM employed at any given time.
$P_{\text{refl}}$ and $P_{\text{trans}}$ values were computed
for various energies $E>V_0$, to a convergence error of $~10^{-4}$,
and in every case found to match the exact
analytical results\cite{ahmed93} to within the same error.
A density plot of the CPWM bipolar components for a
typical solution ($E=450 \, \text{cm}^{-1}\approx0.002$~a.u.)
is presented in Fig.~\ref{ebmanclassfig}. For most energies $E$,
these calculations are roughly as efficient as the constant velocity
calculations described below. However, the required propagation
time does indeed increase rapidly as $E \ra V_0$, as expected.
In this limit, the classical trajectory bipolar decomposition
also becomes unstable; the small central peaks evident
in Fig.~\ref{ebmanclassfig} become increasingly pronounced, eventually
developing into singularities.

\begin{figure}
\includegraphics[scale=0.8]{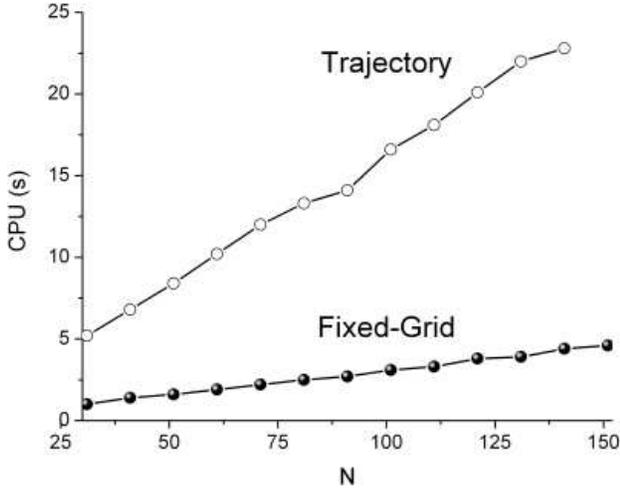}
        \caption{CPU time vs. no. of grid points, $N$, required to propagate
         trajectory (empty circles) and fixed-grid (filled circles)
         implementations of the constant velocity
         CPWM bipolar decomposition to a final system time of
         $t_{\text{max}}=10,000$ a.u. The trajectory and fixed grid time-steps
         used were $\Dlt = 1$ a.u. and $\Dlt = 0.1$ a.u., respectively.}
        \label{cpufig}
\end{figure}

The second study is a detailed convergence and efficiency comparison
between the trajectory and fixed-grid implementations of the constant
velocity method (Sec.~\ref{algorithmic}). This study also serves as
a benchmark for establishing the numerical parameter values needed
to achieve a $10^{-4}$ accuracy level. Both schemes were
applied to a calculation of the $E=V_0$ stationary state, i.e.
the classical ``worst-case scenario,'' for a varying number of
grid points, $N$. A time-step of $\Dlt=1$ a.u. was used for
the trajectory calculation, and $\Dlt=0.1$ a.u. in the fixed-grid case.
Both calculations were propagated to time $t_{\text{max}}=10,000$ a.u.
These parameters were sufficiently converged as to ensure that their
contribution to the total numerical error is insignificant
(i.e. $~10^{-8}$ or less).
In Fig.~\ref{cpufig}, the resultant CPU times on a 2.60 GHz Pentium
computer are presented. Note that despite the ten-fold increase in
the number of time-steps for the fixed-grid case, the total CPU time
is still markedly faster for all grid sizes considered. This is
due both to the search operation (to find the nearest opposite
LM trajectories) and the interpolation matrix inversions
required at each time-step by the trajectory scheme.

\begin{figure}[t!]
\includegraphics[scale=0.8]{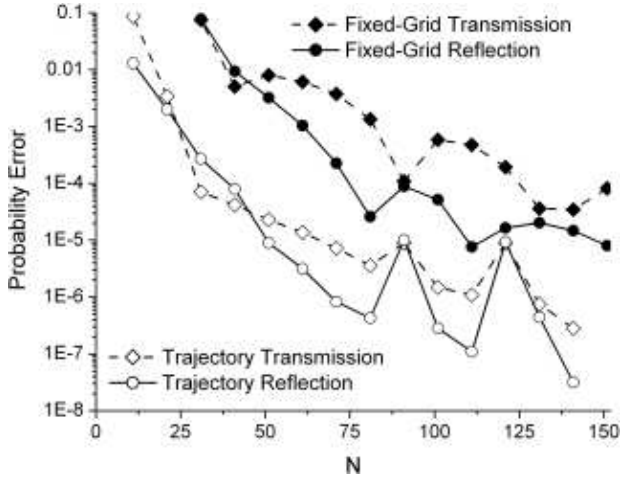}
        \caption{Errors in computed reflection probabilities $P_{\text{refl}}$
                 (solid lines) and transmission probabilities $P_{\text{trans}}$
                 (dashed lines) vs. no. of grid points $N$ for the
                 $E=400\, \text{cm}^{-1}$ stationary state of the
                 $V_0=400\, \text{cm}^{-1}$ Eckart barrier system, as obtained
                 using trajectory and fixed-grid implementations of the constant
                 velocity CPWM bipolar decomposition. Errors are relative
                 to known analytical results (Ref. [47]).}
        \label{errorfig}
\end{figure}

In Fig.~\ref{errorfig}, the log of the computed $P_{\text{refl}}$ and
$P_{\text{trans}}$ errors (relative to known analytical
values\cite{ahmed93})
are plotted versus $N$. Across $N$, the trajectory-based results
are seen to be much more accurate than the fixed-grid results, although
the latter are substantially improved by increasing the density of
grid points. For example, in order to achieve $10^{-4}$ accuracy for
both $P_{\text{refl}}$ and $P_{\text{trans}}$, only $N=25$
trajectories were needed, whereas $N=91$ fixed-grid points were
required. This is consistent with the fixed-grid method requiring
explicit spatial differentiation, introducing an additional source
of numerical error. Note that for this comparison, the CPU times are
comparable---4.3 seconds and 2.7 seconds, respectively. The
fixed-grid calculation is slightly faster, although this situation
would be reversed for more accurate calculations (e.g. $10^{-6}$)
and/or higher dimensionalities. The performance of both methods is
in any event remarkable, with the large-$N$ trajectory case
representative of the most accurate trajectory-based quantum
scattering calculations performed to date.  Further improvements are
also possible, however, as discussed in Sec.~\ref{lowlevel}.

\begin{figure}
\includegraphics[scale=0.8]{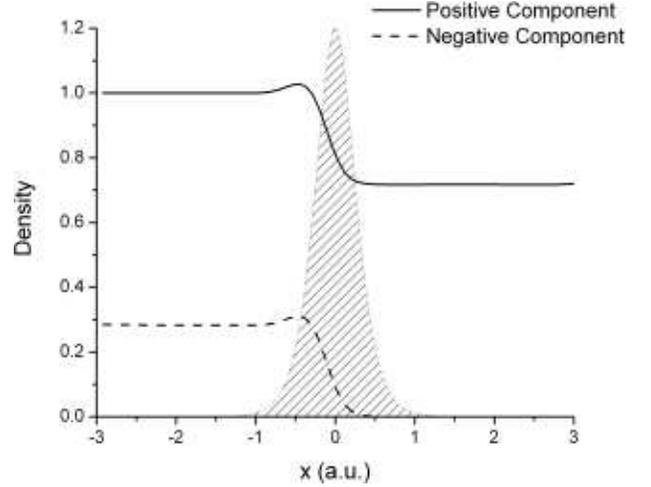}
        \caption{Converged positive component density $\rho_+(x)$ (solid line) and
                 negative component density $\rho_-(x)$ (dashed line) for the
                 $E=400\, \text{cm}^{-1}$ stationary state of the
                 $V_0=400\, \text{cm}^{-1}$ Eckart barrier system, as obtained
                 using the constant velocity trajectory CPWM bipolar decomposition.
                 Shaded area indicates the Eckart potential $V(x)$.}
        \label{ebmanfig}
\end{figure}

\begin{figure}
\includegraphics[scale=0.8]{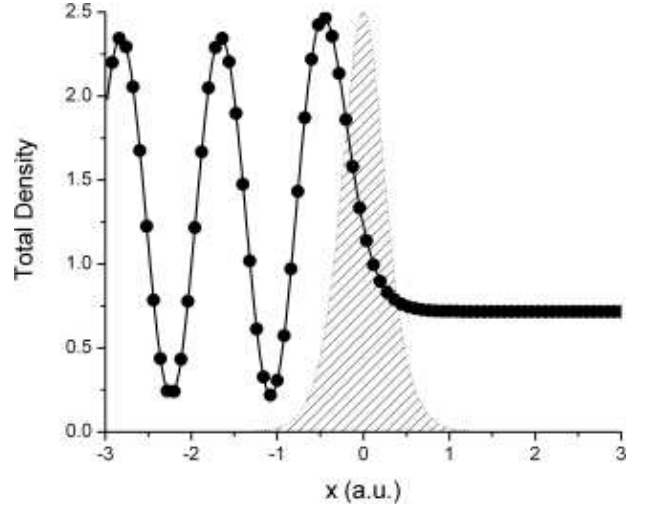}
        \caption{Total wavefunction density $\rho(x)$ for the
                 $E=400\, \text{cm}^{-1}$ stationary state of the
                 $V_0=400\, \text{cm}^{-1}$ Eckart barrier system, as obtained
                 using the constant velocity trajectory CPWM bipolar decomposition
                 (filled circles), and compared with exact analytical
                 results (solid line).
                 Shaded area indicates the Eckart potential $V(x)$.}
        \label{ebdenfig}
\end{figure}

The final study was a constant velocity trajectory calculation
of $P_{\text{refl}}$ and $P_{\text{trans}}$
over a wide range of $E$ values---including those above, below,
and at the barrier maximum, $E=V_0$.
The parameters $\Dlt = 10$ a.u., $t_{\text{max}}=10,000$ a.u.,
and $N=31$ were chosen to correspond to a target accuracy of
$~10^{-4}$. The resulting stationary state solution
for the $E=V_0$ case is displayed in Figs.~\ref{ebmanfig}
and~\ref{ebdenfig}. Note the contrast between the smooth,
slowly-varying CPWM bipolar densities $\rho_\pm$ in Fig.~\ref{ebmanfig}
vs. the oscillatory total density $\rho$ of Fig.~\ref{ebdenfig}.
Nevertheless, as indicated in the latter figure, the numerically
reconstructed total density agrees with the
analytical result\cite{ahmed93}
to within the target accuracy of $10^{-4}$ at all grid points.
Note also that $\rho_+$ and $\rho_-$ are identical
apart from a constant, as predicted in Sec.~\ref{additionaluni}.

\begin{figure}
\includegraphics[scale=0.8]{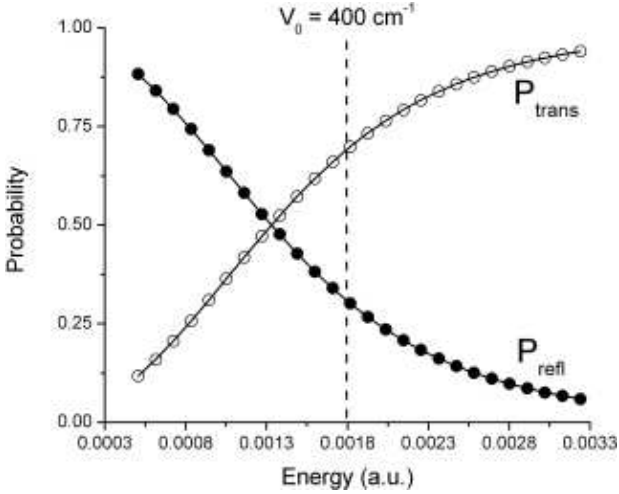}
        \caption{Computed reflection probabilities $P_{\text{refl}}$
                 (filled circles) and transmission probabilities $P_{\text{trans}}$
                 (empty circles) vs energy $E$ for the $V_0=400\, \text{cm}^{-1}$
                  Eckart barrier system, as obtained
                  using the constant velocity trajectory CPWM bipolar decomposition,
                  and compared with exact analytical results (solid line).}
        \label{ebtransfig}
\end{figure}

For all 26 energy values considered,
the resulting constant velocity bipolar decompositions are
qualitatively similar to those presented here for $E = V_0$.
Indeed, in no respect do these calculations seem to depend
sensitively on the value of $E$, unlike the classical
trajectory case. Consequently, the same numerical parameters
as described above were used for all energies. The resulting
computed $P_{\text{refl}}$ and $P_{\text{trans}}$ values are
presented in Fig.~\ref{ebtransfig} and compared with
analytical values. Once again, all errors are found to be
smaller than $10^{-4}$.

\subsection{The Square Barrier}
\label{squarebarrier}

Although the primary focus of this paper is continuous potentials,
it is worth emphasizing that the methods presented here
may be applied to discontinuous potentials as well.
As a case in point, we reexamine the square barrier system
introduced in paper~II. Note that the
classical trajectory scheme for $E>V_0$ would yield results
identical to those of paper~II, as is easily verified
from \eq{Pdotgwf}. Instead, we focus on the constant velocity
trajectory scheme, which can be applied directly
without any algorithmic modification.

For the following square barrier study, we used a barrier height of
$V_0=400\, \text{cm}^{-1}$, and barrier edges of $x_1= -1$ and $x_2 = 1$,
respectively. As in Sec.~\ref{eckart}, $N=31$ initial grid points
were distributed uniformly throughout the interaction region,
and the time-step was defined as $\Dlt = 10$ a.u.
The energy was taken to be $E=450\,\text{cm}^{-1}\approx0.002$ a.u.,
i.e. slightly above-barrier. The propagation was continued
until the computed $P_{\text{refl}}$ and $P_{\text{trans}}$
values were both converged to less than $10^{-4}$, which was
found to require approximately 1000 time-steps.

\begin{figure}
\includegraphics[scale=0.8]{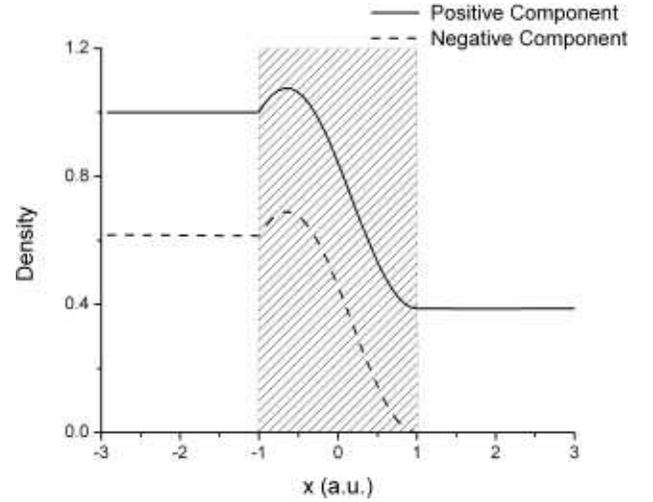}
        \caption{Converged positive component density $\rho_+(x)$ (solid line) and
                 negative component density $\rho_-(x)$ (dashed line) for the
                 $E=450\, \text{cm}^{-1}$ stationary state of the
                 $V_0=400\, \text{cm}^{-1}$ square barrier system, as obtained
                 using the constant velocity trajectory CPWM bipolar decomposition.
                 Shaded area indicates the square barrier potential $V(x)$}
        \label{sbmanfig}
\end{figure}

Figure~\ref{sbmanfig} is a density plot of the converged
CPWM bipolar component solutions. As in the Eckart case, the
results are well-behaved and interference-free throughout,
although there is a discontinuity in the first spatial
derivative at the step edges (but not for the total $\Psi$ itself).
For the total density $\rho$, the computed
values and the analytical results are once again in
agreement to $10^{-4}$ or better, at all grid points.

\begin{figure}
\includegraphics[scale=0.8]{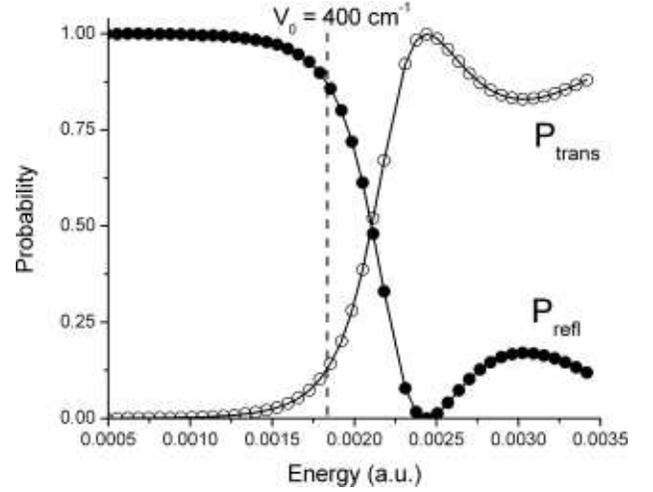}
        \caption{Computed reflection probabilities $P_{\text{refl}}$
                 (filled circles) and transmission probabilities $P_{\text{trans}}$
                 (empty circles) vs energy $E$ for the $V_0=400\, \text{cm}^{-1}$
                  square barrier system, as obtained
                  using the constant velocity trajectory CPWM bipolar decomposition,
                  and compared with exact analytical results (solid line).}
        \label{sbtransfig}
\end{figure}

The calculation described above was repeated for a total of 45
different energy values. The resultant converged
$P_{\text{refl}}$ and $P_{\text{trans}}$ values are presented
in Fig.~\ref{sbtransfig}, as are the exact analytical results.
Relative to the latter, all computed probability errors are
found to be less than $10^{-4}$.

\subsection{The Uphill Ramp}
\label{uphill}

The next application considered is an asymptotically asymmetric
system---the continuous ``uphill ramp''\cite{flugge}
defined via
\eb
     V(x) = {V_0 \over 2} \sof{1+\tanh\of{{x \over 2 \alpha}}},
\ee
with the parameters $V_0=400\, \text{cm}^{-1}$ and  $\alpha=0.2$ a.u.
This is another scattering system that is analytically
soluble.\cite{flugge}
Note that $\lim_{x \ra -\infty} V(x) = 0$ and
$\lim_{x \ra \infty} V(x) = V_0$, as presumed in Sec.~\ref{asymmetric}.

\begin{figure}
\includegraphics[scale=0.8]{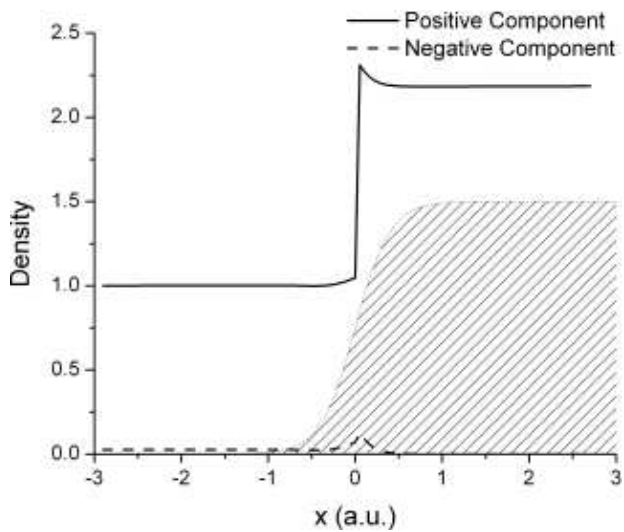}
        \caption{Converged positive component density $\rho_+(x)$ (solid line) and
                 negative component density $\rho_-(x)$ (dashed line) for the
                 $E=500\, \text{cm}^{-1}$ stationary state of the
                 $V_0=400\, \text{cm}^{-1}$ uphill ramp system, as obtained
                 using the constant velocity trajectory CPWM bipolar decomposition,
                 modified for asymptotically asymmetric potentials. The dividing
                 point is $x_0=0$.
                 Shaded area indicates the uphill ramp potential $V(x)$.}
        \label{urmanfig}
\end{figure}

The propagation was performed using the constant velocity
trajectory scheme described in Sec.~\ref{asymmetric}, with
the dividing point chosen to be $x_0=0$. All other computational
parameters were identical to those of the previous examples,
except for the energy, which was chosen to be
$E=500\, \text{cm}^{-1}\approx0.0023$ a.u.
Again, the propagation was continued until both
$P_{\text{refl}}$ and $P_{\text{trans}}$ were
converged to less than $10^{-4}$. Figure~\ref{urmanfig} is a density
plot of the resultant CPWM bipolar component solutions. There is a
discontinuity at $x_0=0$, which is to be expected given the
``step-like'' nature of the join (the same behavior is observed in
paper~II). Importantly, however, this discontinuity does not give
rise to any numerical problems, since spatial derivatives are not
required in the trajectory integration (though one must be a bit
careful with the  interpolation). Away from $x_0$, both density
plots are smooth and well-behaved.

\begin{figure}
\includegraphics[scale=0.8]{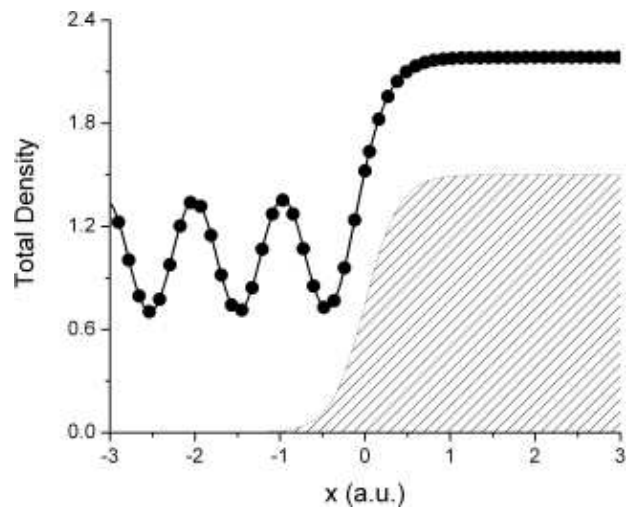}
        \caption{Total wavefunction density $\rho(x)$ for the
                 $E=500\, \text{cm}^{-1}$ stationary state of the
                 $V_0=400\, \text{cm}^{-1}$ uphill ramp system, as obtained
                 using the constant velocity trajectory CPWM bipolar decomposition
                 (filled circles), and compared with exact analytical
                 results (solid line, Ref. [48]).
                 Shaded area indicates the uphill ramp potential $V(x)$.}
        \label{urdenfig}
\end{figure}

Despite the discontinuities in $\rho_\pm$, the computed $\rho$ and its
first derivative are continuous, owing to the \eq{asymbcB} relations.
This is evident in Fig.~\ref{urdenfig}, a density plot for the converged
total $\Psi$. As is clear from the plot, no visible discrepancies
may be observed between the computed and analytically exact solutions.

\subsection{The Double-Gaussian Barrier}
\label{dgaussian}

The previous three example systems serve as useful benchmarks
for testing and evaluating the new bipolar methodologies
under a wide range of conditions. In particular, all three
have known analytical solutions, against which the computed
results may be compared. However, we feel it is also worthwhile
to consider at least one system which has not previously
been solved. One such example, which also presents a qualitatively
new variety of problem, is the symmetric double-Gaussian barrier
potential,
\eb
V(x)=V_0 \left \{ \exp\sof{-\beta \of{x-x_0}^2}
             +\exp\sof{-\beta \of{x+x_0}^2} \right \}.
\ee
For the results presented here, the following
parameters were used: $V_0=400\, \text{cm}^{-1}$; $\beta=9$~a.u.;
$x_0=0.75$~a.u.

\begin{figure}
\includegraphics[scale=0.8]{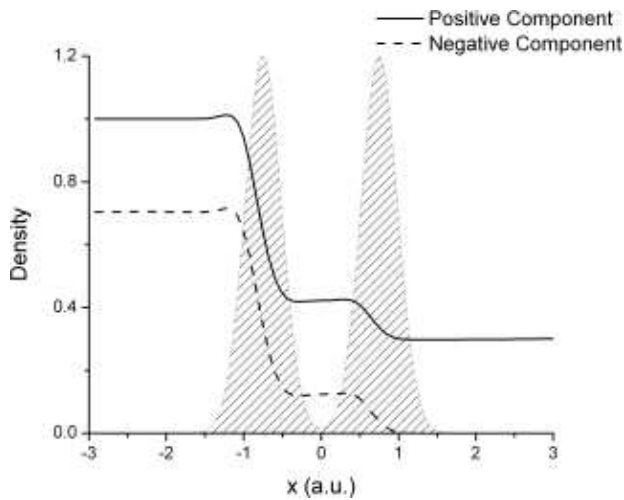}
        \caption{Converged positive component density $\rho_+(x)$ (solid line) and
                 negative component density $\rho_-(x)$ (dashed line) for the
                 $E=400\, \text{cm}^{-1}$ stationary state of the
                 $V_0=400\, \text{cm}^{-1}$ double-Gaussian barrier system, as obtained
                 using the constant velocity trajectory CPWM bipolar decomposition.
                 Shaded area indicates the double-Gaussian potential, $V(x)$.}
        \label{dgmanfig}
\end{figure}

Once again, the constant velocity trajectory method was employed,
with the same computational parameters
as described previously. The energy was chosen to be just at the
barrier height, $E = V_0$. Figure~\ref{dgmanfig} displays the
resulting converged CPWM bipolar densities. As in the previous examples,
these are identical apart from a constant, and are otherwise
smooth and slowly-varying. Note that the double-barrier nature
of the potential gives rise to interesting features in the
bipolar densities not previously observed. In particular,
despite the fact that $V(x)$ is changing in the central well
region between the two barriers, the bipolar densities are
nearly flat across this region, resulting in a well-defined
``reaction intermediate'' state between reactants and
products. Moreover, the density plots enable one to assign
quantitative probability values for the intermediate state.

\begin{figure}
\includegraphics[scale=0.8]{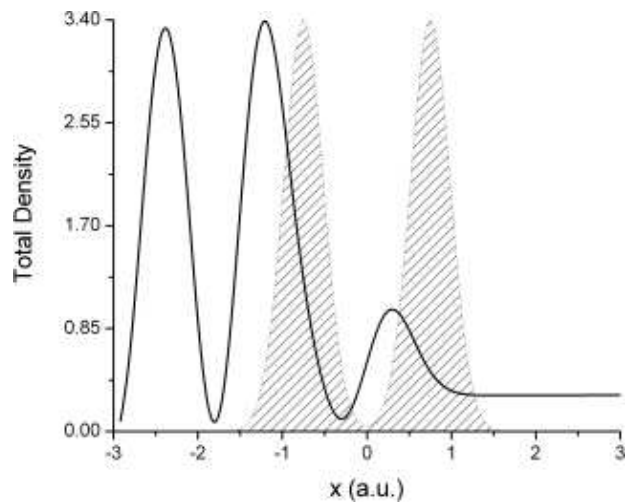}
        \caption{Total wavefunction density $\rho(x)$ (solid line) for the
                 $E=400\, \text{cm}^{-1}$ stationary state of the
                 $V_0=400\, \text{cm}^{-1}$ double-Gaussian barrier system, as obtained
                 using the constant velocity trajectory CPWM bipolar decomposition.
                 Shaded area indicates the double-Gaussian potential, $V(x)$.}
        \label{dgdenfig}
\end{figure}

In contrast, the above interpretation and quantitative
assignments would be very difficult, if not impossible, to
glean directly from $\Psi(x)$ itself. This is evident
from Fig.~\ref{dgdenfig}, a density plot of the total
$\Psi$ for the above double-Gaussian barrier calculation,
obtained via interpolation and superposition of the two converged
bipolar component solutions, $\Ppm$. Note the interference
present both in the reactant region (due to reflected
trajectories) {\em and} the intermediate region. Thus,
not only the intermediate probabilities, but also the
reflection probability, are difficult to read directly
from such a plot.

\section{SUMMARY AND CONCLUSIONS}
\label{conclusion}

Scattering applications are of paramount importance for chemical
reactions, because all reactions may be regarded as
scattering events. From a theoretical exact quantum perspective,
therefore, multichannel scattering theory,\cite{taylor}
both time-dependent and time-independent, will always
play an essential role. At the
same time however, trajectory-based methods also bring much to
bear on dynamics, providing great insight into
reactive processes, vis-a-vis the determination of which
trajectories make it past the barrier to products
vs. those that do not. Quantum trajectory methods (QTMs) therefore
exhibit great potential promise as a chemical dynamics tool,
combining a trajectory-based description with exact quantum
dynamics. However, {\em all} reactive systems exhibit interference
between the incident and reflected (non-reactive) waves, thus
causing numerical instability problems for conventional unipolar
QTMs. CPWM bipolar decompositions offer a natural means of alleviating
this interference difficulty, by splitting the reactant region
$\Psi$ into incident $\Psi_+$ and reflected $\Psi_-$ components,
neither of which exhibits interference on its own.
Moreover, this splitting can be extended through the interaction
region over to the product region, by which point
$\Psi_+$ has transformed smoothly into the transmitted wave,
and $\Psi_-$ has damped to zero.

As discussed in Sec.~\ref{theory} and the Appendices, the CPWM
approach borrows conceptually from semiclassical scattering methods.
Indeed, for the first bipolar decomposition considered
(B, Sec.~\ref{classicaltraj}) the bipolar trajectories are
simply equal to the classical trajectories themselves, and
the correspondence principle is satisfied in precisely the
usual WKB limit, $V' \ra 0$. Three features, however,
contribute to render the present approach fundamentally
different from basic WKB theory: (1) time-dependent formulation;
(2) coupling between $\Psi_+$ and $\Psi_-$; (3) universal,
local reflection and transmission formulae (see also paper~II).
Point (3) is what determines point (2), i.e. if only local
transmission were considered without reflection, then the coupling
would vanish and the basic WKB solutions would result.  The
combination of (1) and (3) provides a local physical
understanding related to the ray optics picture in
electromagnetic theory, and also gives rise to useful flux
relations and numerical algorithms.
Regarding the second bipolar decomposition considered,
i.e. the F, constant velocity scheme, this was motivated
by practical concerns, but also by
an alternative ray optics description (Sec.~\ref{constanttraj}).
This can be related to the semiclassical approach
of Fr\"oman and Fr\"oman, and in that context,
also satisfies a generalized kind of correspondence
principle.

Note that neither set of evolution
equations [\eqs{Pdotgwf}{Pdotunigwf}] involves
a quantum potential; instead, all quantum effects manifest
through $\Ppm$ coupling. In both cases, the trajectories
themselves are not ``context-sensitive,''\cite{wyatt} in
that they may be computed independently of the $\Ppm$
evolution. Moreover, no spatial differentiation of the
wavefunction is required, although there may be situations
where explicit calculation of one spatial
derivative is numerically advantageous (Sec.~\ref{eckart}).
Several other numerical modifications have also been
introduced for convenience (Sec.~\ref{algorithmic}), resulting
in a shift from an exact time-dependent \shro\ interpretation
of the dynamics [wherein the exact stationary state is ``revealed''
over time (paper~II)] to what may be regarded as
more of a relaxation approach. Be that as it may,
the resulting algorithms offer a remarkably simple,
efficient, and accurate means of performing reactive scattering
calculations of all kinds in 1D (Sec.~\ref{results}). Indeed,
the time-steps required for the benchmark molecule-like systems
considered here are orders of magnitude larger than for typical
fixed-grid calculations performed at a comparable level of accuracy.
Moreover, the converged bipolar solution density plots render
the determination of global reflection and transmission
probabilities, as well as probabilities for reaction
intermediate states, quite straightforward.

In future publications we will continue to generalize the
methodology described here and in paper~II, for the type of
multidimensional time-dependent wavepacket dynamics relevant to
real chemical physics applications. As additional motivation for
the present work, we now sketch how this might be achieved.
First, it is necessary to generalize the stationary state
results of this paper and paper~II for arbitrary time-evolving
wavepackets. This is done initially for the discontinuous
step potential, and then generalized for arbitrary continuous
potentials in a manner similar to Appendix A. In fact, much of the
groundwork is already laid, in that the time-dependent framework
has already been introduced.

The generalization to multidimensional systems is less straightforward
but can certainly be achieved (such calculations have already been
performed, as will be reported in a future publication).
Conceptually at least,
many direct chemical reactions can be described using a single scattering
reaction coordinate, plus additional ``bound'' coordinates.
It is natural to consider applying the current CPWM bifurcation to
the former, and the paper~I bifurcation to each of the latter.
However, the total number of wavefunction components
would then be $2^D$ where $D$ is the number of degrees of freedom.
On the other hand, for most time-dependent wavepacket calculations,
node formation in $\Psi$ is associated primarily with the reaction
coordinate itself, due to wavepacket reflection off of the
reaction profile barrier. Thus, a natural approach would be to
bifurcate {\em only} along the reaction coordinate. Only two
component wavefunctions result, regardless of $D$.
It remains to be seen whether such a procedure will render QTM
calculations possible for actual molecular systems. Nevertheless,
it seems very likely that some such bipolar or multipolar approach
will go a long way towards ameliorating the infamous node problem,
which has thus far severely limited the effectiveness of QTMs
in the molecular arena.

\begin{acknowledgments}

This work was supported by awards from The Welch Foundation
(D-1523) and Research Corporation. The authors would like to acknowledge
Robert E. Wyatt and Eric R. Bittner for many stimulating discussions.
David J. Tannor and John C. Tully  are also acknowledged.
Jason McAfee is also acknowledged for his aid in converting this manuscript to an electronic format suitable for the arXiv preprint server.

\end{acknowledgments}
%
\appendix

\section{Derivation of bipolar time-evolution equations}

Following the notation of paper~II, Sec. II D 3, we presume a
{\em discontinuous} potential, for which $x_k$
($k=1,2,\ldots,l$) denote the locations of the $l$
discontinuities. The $x_k$ divide configuration space
into $l+1$ regions, labeled $0\le k \le l$. In each
region $k$, the potential has a different constant
value, $V_k$. The discontinuous system described above
may be used to model any {\em continuous} potential
system, $V(x)$, by defining $V_k = V(x_k^m)$ [where
$x_k^m = (x_k + x_{k+1})/2$ is the region midpoint] and
taking the limit that $(x_{k+1} - x_k) \ra 0$ for all $k$.

As the derivation is a time-dependent one, it is convenient
to introduce a small (ultimately differential) time increment
$\Dlt$, which is then used to determine the $x_k$ as follows.
Associated with each region $k$ is the positive classical
momentum value, $p_k = \sqrt{2 m \sof{E-V_k}}$ [i.e.
$p_+^\sc(x_k^m)$]. The $x_k$ are chosen such that
a particle moving with momentum $p_k$ would traverse
the region $k$ in time $\Dlt$, i.e.
$(x_{k+1}- x_k) = \Dlt (p_k/m)$. Consider a trajectory which
at $t=0$, is located at the $k$'th region midpoint, $x_k^m$,
heading to the right with momentum $p_k$. This is clearly
a positive LM trajectory, carrying  a contribution of the
component wavefunction $\Psi_{+k} = \Psi_+(x=x_k^m,t=0)$
(Fig.~\ref{derivefig}). It is presumed that there are also
negative LM trajectories along which $\Psi_-$ is propagated,
but for now the emphasis is on $\Psi_+$.

From $t= 0$ to $t = \Dlt/2$, the positive LM trajectory
travels from $x=x_k^m$ to the discontinuity at $x=x_{k+1}$.
As per paper~II, the propagation is that of a plane wave, i.e.
\ea{
     \Psi_{+k} & \ra & \Psi_+(x_{k+1},\Dlt/2)\nonumber\\
                   & & =  \Psi_{+k} \exp\of{{i \Dlt \over 2 \hbar}
                   \sof{{p_k^2 \over 2m} - V_k}} \nonumber \\
                   &  & = \Psi_{+k} \exp\of{{i \Dlt \over 2 \hbar}
                   \sof{E - 2 V_k}}. \label{firsthalf}}
At this point, the trajectory splits into two: one that
continues in the forward direction along the positive LM,
transmitting into region $k+1$ with momentum $p_{k+1}$;
the other reflected backwards along the negative LM with
momentum $-p_k$ (Fig.~\ref{derivefig}).  According to
paper~II Eqs.~(17) and~(18), the trajectory bifurcation
introduces a factor of $2 p_k/(p_k+p_{k+1})$ into the
the transmitted $\Psi_+$ wave, and
$(p_k-p_{k+1})/(p_k+p_{k+1})$ into the reflected
$\Psi_-$ wave.

During the remaining time evolution from $t= \Dlt/2$ to
$t=\Dlt$, the $\Psi_+$ trajectory moves to $x_{k+1}^m$
[the midpoint of the adjacent $(k+1)$'th region], resulting
in an additional phase factor analogous to that of
\eq{firsthalf}. The final result is
\ea{
    && \Psi_{+k} \ra
                   \Psi_+(x_{k+1}^m,\Dlt) \\
                  &&   = \Psi_{+k} \exp\of{{i \Dlt \over \hbar}
                   \sof{E -  V_k - V_{k+1}}}
               \of{{2 p_k \over p_k+p_{k+1}}}.\nonumber\label{secondhalftrans}}
The reflected trajectory, meanwhile, moves back to the original
location at $x_k^m$, resulting in the following for $\Psi_-$:
\ea{
     &&\Psi_{+k} \ra
                   \Psi_-(x_k^m,\Dlt)\\
                   &  & =\Psi_{+k} \exp\of{{i \Dlt \over \hbar}
                   \sof{E -  2 V_k}}
               \of{{p_k-p_{k+1} \over p_k+p_{k+1}}}.\nonumber \label{secondhalfrefl}}

We thus find that $\Psi_{+k}$ at time $t=0$ contributes to both
$\Psi_{+(k+1)}$ and $\Psi_{-k}$ at time $t= \Dlt$. However, these must be
combined with similar contributions from the initial $\Psi_{-(k+1)}$ in order to
determine the total final $\Psi_{+(k+1)}$ and $\Psi_{-k}$ (Fig.~\ref{derivefig}).
Following an analysis similar to the above, the negative LM contributions are
easily shown to be the following:
\ea{
     &&\Psi_{-(k+1)} \ra
                   \Psi_+(x_{k+1}^m,\Dlt)  \label{secondhalfreflmin} \\
                  && = - \Psi_{-(k+1)} \exp\of{{i \Dlt \over \hbar}
                   \sof{E -  2 V_{k+1}}}
               \of{{p_k-p_{k+1} \over p_k+p_{k+1}}}.\nonumber\\
     &&\Psi_{-(k+1)} \ra
                   \Psi_-(x_k^m,\Dlt)\label{secondhalftransmin}\\
                   & & = \Psi_{-(k+1)} \exp\of{{i \Dlt \over \hbar}
                   \sof{E -  V_k - V_{k+1}}}
               \of{{2 p_{k+1} \over p_k+p_{k+1}}}.\nonumber }

By adding \eqs{secondhalftrans}{secondhalfreflmin}, we obtain the final
expression for $\Psi_+(x_{k+1}^m,\Dlt)$. Subtracting $\Psi_{+k}$, dividing
by $\Dlt$, and taking the limit $\Dlt \ra 0$ yields the total (hydrodynamic)
time derivative, $d \Psi_+ / dt$. We thus obtain
\ea{
   &&\lim_{\Dlt \ra 0} \sof{{\Psi_+(x_{k+1}^m,\Dlt) - \Psi_{+k} \over \Dlt}}  =  \label{gwfpart}\nonumber\\
   &&\lim_{\Dlt \ra 0} \Bigg\{\Bigg.\bigg[{1 \over \Dlt} \of{{p_k-p_{k+1} \over p_k+p_{k+1}}}
   + {i\over \hbar}\of{E - V_k - V_{k+1}}\nonumber\\
   &&  \times \of{{2 p_k \over p_k+p_{k+1}}}\Bigg.\Bigg] \Psi_{+k}
    - \sof{ {1 \over \Dlt} + {i\over \hbar}\of{E - 2 V_{k+1}}}\nonumber\\
    && \times\of{{p_k-p_{k+1} \over p_k+p_{k+1}}} \Psi_{-(k+1)}\Bigg.\Bigg\}.}
Recall that $\Dlt = (x_{k+1}-x_k) m/p_k$. In the small $\Dlt$ limit,
$V_{k+1} \ra V_k$, $p_{k+1} \ra p_k$, and
$-(p_k-p_{k+1})/(x_{k+1}-x_k) \ra p'(x) = \partial p(x)/ \partial x$.
Substituting into \eq{gwfpart}, and replacing $k$ subscripts with functions
of $x=x_k$ yields:
\eb
     {d \Psi_+ \over dt} = \sof{ - {p' \over 2m} + {i \over \hbar} \of{E-2V}} \Psi_+
     + {p' \over 2m} \Psi_- \label{Pplusdotgwf}
\ee
A similar analysis applied to \eqs{secondhalfrefl}{secondhalftransmin} yields:
\eb
     {d \Psi_- \over dt} = \sof{ {p' \over 2m} + {i \over \hbar} \of{E-2V}} \Psi_-
     - {p' \over 2m} \Psi_+ \label{Pminusdotgwf}
\ee

\section{Proof that bipolar stationary solutions satisfy time-independent
Schr\"odinger equation}

Consider the stationary solutions of \eq{Pdot}, i.e.
$\partial \Ppm /\partial t = -(i/\hbar) E \Ppm$. Substituting in these
expressions for the time derivatives, and rewriting to obtain expressions
for the spatial derivatives, yields:
\eb
     \Ppm' = \of{-{p' \over 2p} \pm {i p \over \hbar}}\Ppm + {p' \over 2p} \Pmp
     \label{berrymount}
\ee
These equations are identical to those of the bipolar time-independent stationary
state decomposition described in \Ref{berry72}. Adding the ${\Psi_+}'$ and ${\Psi_-}'$
equations together results in
\eb
     \Psi' = {i p \over  \hbar} \of{\Psi_+ - \Psi_-}. \label{BPprime}
\ee
Applying spatial differentiation and substituting \eq{berrymount} into the resulting
right hand side yields $\Psi'' = - (p^2/\hbar^2) \Psi$. Substitution into the
\shro\ equation then results in
\eb
- {\hbar^2 \over 2m} \Psi'' + V \Psi = {p^2 \over 2m} \Psi + V \Psi = E \Psi.
\ee
The stationary solutions of \eq{Pdot} are therefore
consistent with the time-independent \shro\ equation.

%
%

%
%
%



%
%


\begin{thebibliography}{10}

\bibitem{poirier04bohmI}
B. Poirier, J. Chem. Phys. {\bf 121},  4501  (2004).

\bibitem{poirier05bohmII}
B. Poirier, J. Chem. Phys.  , (in press).

\bibitem{babyuk04}
D. Babyuk and R.~E. Wyatt, J. Chem. Phys. {\bf 121},  9230  (2004).

\bibitem{wyatt}
R.~E. Wyatt, {\em Quantum Dynamics with Trajectories: Introduction to Quantum
  Hydrodynamics} (Springer, New York, 2005).

\bibitem{lopreore99}
C.~L. Lopreore and R.~E. Wyatt, Phys. Rev. Lett. {\bf 82},  5190  (1999).

\bibitem{mayor99}
F.~S. Mayor, A. Askar, and H.~A. Rabitz, J. Chem. Phys. {\bf 111},  2423
  (1999).

\bibitem{wyatt99}
R.~E. Wyatt, Chem. Phys. Lett. {\bf 313},  189  (1999).

\bibitem{shalashilin00}
D.~V. Shalashilin and M.~S. Child, J. Chem. Phys. {\bf 113},  10028  (2000).

\bibitem{wyatt01b}
R.~E. Wyatt and E.~R. Bittner, J. Chem. Phys. {\bf 113},  8898  (2001).

\bibitem{wyatt01c}
R.~E. Wyatt and K. Na, Phys. Rev. E {\bf 65},  016702  (2001).

\bibitem{burghardt01b}
I. Burghardt and L.~S. Cederbaum, J. Chem. Phys. {\bf 115},  10312  (2001).

\bibitem{bittner02b}
E.~R. Bittner, J.~B. Maddox, and I. Burghardt, Int. J. Quantum Chem. {\bf 89},
  313  (2002).

\bibitem{hughes03}
K.~H. Hughes and R.~E. Wyatt, Phys. Chem. Chem. Phys. {\bf 5},  3905  (2003).

\bibitem{madelung26}
E. Madelung, Z. Phys. {\bf 40},  322  (1926).

\bibitem{vanvleck28}
J.~H. {van Vleck}, Proc. Natl. Acad. Sci. U.S.A. {\bf 14},  178  (1928).

\bibitem{bohm52a}
D. Bohm, Phys. Rev. {\bf 85},  166  (1952).

\bibitem{bohm52b}
D. Bohm, Phys. Rev. {\bf 85},  180  (1952).

\bibitem{takabayasi54}
T. Takabayasi, Prog. Theor. Phys. {\bf 11},  341  (1954).

\bibitem{holland}
P.~R. Holland, {\em The Quantum Theory of Motion} (Cambridge University Press,
  Cambridge, 1993).

\bibitem{floyd94}
E.~R. Floyd, Physics Essays {\bf 7},  135  (1994).

\bibitem{brown02}
M.~R. Brown, arXiv:quant-ph/0102102  (2002).

\bibitem{taylor}
J.~R. Taylor, {\em Scattering Theory} (John Wiley \& Sons, Inc., New York, NY,
  1972).

\bibitem{heading}
J. Heading, {\em An Introduction to Phase-integral Methods} (Methuen, London,
  1962).

\bibitem{froman}
N. Fr{\"o}man and P.~O. Fr{\"o}man, {\em JWKB Approximation} (North-Holland,
  Amsterdam, 1965).

\bibitem{berry72}
M.~V. Berry and K.~V. Mount, Rep. Prog. Phys. {\bf 35},  315  (1972).

\bibitem{keller60}
J.~B. Keller and S.~I. Rubinow, Ann. Phys. {\bf 9},  24  (1960).

\bibitem{maslov}
V.~P. Maslov, {\em Th{\'e}orie des Perturbations et M{\'e}thodes Asymptotiques}
  (Dunod, Paris, 1972).

\bibitem{littlejohn92}
R.~G. Littlejohn, J. Stat. Phys. {\bf 68},  7  (1992).

\bibitem{child}
M.~S. Child, {\em Molecular Collision Theory} (Dover, New York, 1996).

\bibitem{poirier03capII}
B. Poirier and T. {Carrington, Jr.}, J. Chem. Phys. {\bf 119},  77  (2003).

\bibitem{bremmer51}
H. Bremmer, Commun. Pure Appl. Math. {\bf 4},  105  (1951).

\bibitem{poirier03capI}
B. Poirier and T. {Carrington, Jr.}, J. Chem. Phys. {\bf 118},  17  (2003).

\bibitem{jackson}
J.~D. Jackson, {\em Classical Electrodynamics}, 2nd  ed. (John Wiley \& Sons,
  New York, 1975).

\bibitem{brillouin14}
L. Brillouin, Ann. Phys. {\bf 44},  177  (1914).

\bibitem{hirschfelder74}
J.~O. Hirschfelder, A.~C. Christoph, and W.~E. Palke, J. Chem. Phys. {\bf 61},
  5435  (1974).

\bibitem{complexscaling78}
see~for example, Int. J. Quant. Chem. {\bf 14},    (1978), special issue.

\bibitem{reinhardt82}
W.~P. Reinhardt, Ann. Rev. Phys. Chem. {\bf 33},  223  (1982).

\bibitem{ryaboy94}
V. Ryaboy, N. Moiseyev, V.~A. Mandelshtam, and H.~S. Taylor, J. Chem. Phys.
  {\bf 101},  5677  (1994).

\bibitem{jolicard85}
G. Jolicard and E.~J. Austin, Chem. Phys. Lett. {\bf 121},  106  (1985).

\bibitem{seideman92a}
T. Seideman and W.~H. Miller, J. Chem. Phys. {\bf 96},  4412  (1992).

\bibitem{riss93}
U.~V. Riss and H.-D. Meyer, J. Phys. B: At. Mol. Phys. {\bf 26},  4503  (1993).

\bibitem{muga04}
J.~G. Muga, J.~P. Palao, B. Navarro, and I.~L. Egusquiza, Phys. Rep. {\bf 395},
   357  (2004).

\bibitem{phillips}
G.~M. Phillips and P.~J. Taylor, {\em Theory and Applications of Numerical
  Analysis} (Academic Press, New York, 1996).

\bibitem{press}
W.~H. {Press \em et al},  in {\em Numerical Recipes}, 1st  ed. (Cambridge
  University Press, Cambridge, England, 1989).

\bibitem{fornberg88}
B. Fornberg, Math. Comp. {\bf 51},  699  (1988).

\bibitem{eckart30}
C. Eckart, Phys. Rev. {\bf 35},  1303  (1930).

\bibitem{ahmed93}
Z. Ahmed, Phys. Rev. A {\bf 47},  4761  (1993).

\bibitem{flugge}
S. Flugge, {\em Practical Quantum Mechanics} (Springer-Verlag, New York, 1971),
  Vol.~1.

\end{thebibliography}
\end{document}